\RequirePackage{fix-cm}
\documentclass[pdftex,twocolumn,epjc3,final]{svjour3}  
\smartqed  % flush right qed marks, e.g. at end of proof
\RequirePackage{graphicx}
\RequirePackage{esvect}
\RequirePackage{lineno}
\RequirePackage[T1]{fontenc}
\journalname{Eur. Phys. J. A}
\begin{document}

\title{Beam charge asymmetries for deeply virtual Compton scattering off the proton}

\titlerunning{pDVCS-BCA@CLAS12}   % if too long for running head

\author{
V.~Burkert\thanksref{addr2} 
\and
L.~Elouadrhiri\thanksref{addr2} 
\and
F.-X.~Girod\thanksref{addr3} 
\and
S.~Niccolai\thanksref{addr1} 
\and
E.~Voutier\thanksref{addr1,Cp} 
\and
A.~Afanasev\thanksref{addr23} 
\and
L.~Barion\thanksref{addr11} 
\and
M.~Battaglieri\thanksref{addr2,addr111} 
\and
J.C.~Bernauer\thanksref{addr20,addr22} 
\and
A.~Bianconi\thanksref{addr7,addr8} 
\and
R.~Capobianco\thanksref{addr3}  
\and
M.~Caudron\thanksref{addr1} 
\and
L.~Causse\thanksref{addr1} 
\and
P.~Chatagnon\thanksref{addr1} 
\and
T.~Chetry\thanksref{addr16} 
\and
G.~Ciullo\thanksref{addr12,addr11} 
\and
P.L.~Cole\thanksref{addr6} 
\and
M.~Contalbrigo\thanksref{addr11} 
\and
G.~Costantini\thanksref{addr7,addr8} 
\and
M.~Defurne\thanksref{addr13} 
\and
A.~Deur\thanksref{addr2} 
\and
S.~Diehl\thanksref{addr3,addr5} 
\and
R.~Dupr\'e\thanksref{addr1} 
\and
M.~Ehrhart\thanksref{addr1} 
\and
I.P.~Fernando\thanksref{addr15} 
\and
A.~Filippi\thanksref{addr21} 
\and
T.~Forest\thanksref{addr19} 
\and
J.~Grames\thanksref{addr2} 
\and
P.~Gueye\thanksref{addr10} 
\and
S.~Habet\thanksref{addr1} 
\and
D.~Higinbotham\thanksref{addr2} 
\and
A.~Hobart\thanksref{addr1} 
\and
C.E.~Hyde\thanksref{addr17} 
\and
K.~Joo\thanksref{addr3} 
\and
A.~Kim\thanksref{addr3} 
\and
V.~Klimenko\thanksref{addr3} 
\and
H.-S.~Ko\thanksref{addr1} 
\and
V.~Kubarovsky\thanksref{addr2} 
\and
M.~Leali\thanksref{addr7,addr8} 
\and
P.~Lenisa\thanksref{addr12,addr11} 
\and
D.~Marchand\thanksref{addr1} 
\and
V.~Mascagna\thanksref{addr7,addr8} 
\and
M.~McCaughan\thanksref{addr2} 
\and
B.~McKinnon\thanksref{addr14} 
\and
A.~Movsisyan\thanksref{addr11} 
\and
C.~Mu\~noz~Camacho\thanksref{addr1} 
\and
L.~Pappalardo\thanksref{addr12,addr11} 
\and
E.~Pasyuk\thanksref{addr2} 
\and
M.~Poelker\thanksref{addr2} 
\and
K.~Price\thanksref{addr1} 
\and
B.~Raue\thanksref{addr4} 
\and
M.~Shabestari\thanksref{addr18} 
\and
R.~Santos\thanksref{addr3} 
\and
V.~Sergeyeva\thanksref{addr1} 
\and
I.~Strakovsky\thanksref{addr23} 
\and
P.~Stoler\thanksref{addr3} 
\and
L.~Venturelli\thanksref{addr7,addr8} 
\and
S.~Zhao\thanksref{addr1} 
\and
Z.W.~Zhao\thanksref{addr9} 
\and
and the CLAS Collaboration 
}

\institute{Universit\'e Paris-Saclay, CNRS/IN2P3, IJCLab, 91405 Orsay, France \label{addr1}
\and
Thomas Jefferson National Accelerator Facility, Newport News, VA 23606, USA \label{addr2}
\and
University of Connecticut, Storrs, CT U-3046, USA \label{addr3}
\and
Florida International University, Miami, FL 33199, USA \label{addr4}
\and
Lamar University, Beaumont, TX 77710, USA \label{addr6}
\and
Universit\`a degli Studi di Brescia, 25121 Brescia, Italy \label{addr7}
\and
University of Virginia, Charlottesville, VA 22904, USA \label{addr15}
\and
Duke University, Durham, NC 27708, USA \label{addr9}
\and
Michigan State University, East Lansing, MI 48824, USA \label{addr10}
\and
Universit\`a di Ferrara, 44121 Ferrara, Italy \label{addr12}
\and
INFN, Sezione di Ferrara, 44122 Ferrara, Italy \label{addr11}
\and
INFN, Sezione di Genova, 16100 Genova, Italy \label{addr111}
\and
Universit\"at Gie\ss en, 35390 Gie\ss en, Germany \label{addr5}
\and
IRFU, CEA, Universit\'e Paris-Saclay, 91191 Gif-sur-Yvette, France \label{addr13}
\and
University of Glasgow, Glasgow G12 8QQ, United Kingdom \label{addr14}
\and
Mississippi State University, Mississippi State, MS 39762, USA \label{addr16}
\and
Old Dominion University, Norfolk, VA 23529, USA \label{addr17}
\and
INFN, Sezione di Pavia, 27100 Pavia, Italy \label{addr8}
\and
University of West Florida, Pensacola, FL 32514, USA \label{addr18}
\and
Idaho State University, Pocatello, ID 83209, USA \label{addr19}
\and
Center for Frontiers in Nuclear Science,Stony Brook University, Stony Brook, NY 11794, USA \label{addr20}
\and
INFN, Sezione di Torino, 10125 Torino, Italy \label{addr21}
\and
Riken BNL Research Center, Upton, NY 11973, USA \label{addr22}
\and
The George Washington University, Washington, DC 20052, USA \label{addr23}
}

\thankstext{Cp}{Contact person: voutier@ijclab.in2p3.fr}

\authorrunning{V.~Burkert {\it et al.}} % if too long for running head

\date{Draft : \today}

\maketitle

%
%----------------------------------------------------------------------------------------------------
%

\begin{abstract}

The unpolarized and polarized Beam Char\-ge Asymmetries (BCAs) of the $\vv{e}^{\pm}p \to e^{\pm}p \gamma$ process off unpolarized hydrogen are discussed. The measurement of BCAs with the CLAS12 spectrometer at the Thomas Jefferson National Accelerator Facility, using polarized positron and electron beams at 10.6~GeV is investigated. This experimental configuration allows to measure azimuthal and $t$-dependences of the unpolarized and polarized BCAs over a large $(x_B,Q^2)$ phase space, providing a direct access to the real part of the Compton Form Factor (CFF) ${\mathcal H}$. Additionally, these measurements confront the Bethe-Heitler dominance hypothesis and eventual effects beyond leading twist. The impact of potential positron beam data on the determination of CFFs is also investigated within a local fitting approach of experimental observables. Positron data are shown to strongly reduce correlations between CFFs and consequently improve significantly the determination of $\Re {\rm e} [\mathcal{H}]$. 


\end{abstract}

%
%----------------------------------------------------------------------------------------------------
%

\hyphenation{mo-del}

%
%----------------------------------------------------------------------------------------------------
%

\section{Introduction}
\label{intro}

The understanding of the structure and dynamics of the nucleon remains a major goal of modern Nuclear Physics despite extensive experimental scrutiny. From the initial measurements of elastic electromagnetic form factors to the accurate determination of parton distributions through deep inelastic scattering, the experiments have increased in statistical and systematic precision thanks to the development of powerful electron beams and detector systems. The availability of high intensity, continuous polarized electron beams with high energy is providing today an unprecedented but still limited insight into the nucleon structure problem.

Over the past two decades, the Generalized Parton Distributions (GPDs)~\cite{Mueller:1998fv} have offered a universal and powerful way to characterize nucleon structure, generalizing and unifying the special cases of form factors and parton distribution functions (see~\cite{Diehl:2003ny,BELITSKY20051} for a review). The GPDs are integrals of the Wigner quantum phase space distribution of partons in the nucleon, describing the distribution of particles with respect to both the position and momentum in a quantum-mechanical system~\cite{Ji:2003ak,Belitsky:2003nz}. They encode the correlation between partons and consequently reveal not only the spatial and momentum densities, but also the correlation between the spatial and momentum distributions, {\it i.e.} how the spatial shape of the nucleon changes when probing partons of different momentum fraction $x$ of the nucleon. The combination of longitudinal and transverse degrees of freedom is responsible for the richness of this framework. The second moments in $x$ of the GPDs are related to form factors that allow us to quantify how the orbital motion of partons in the nucleon contributes to the nucleon spin~\cite{Ji:1996ek}, and how the parton masses and the forces on partons are distributed in the transverse space~\cite{Polyakov:2002yz}, a question of crucial importance for the understanding of the dynamics underlying nucleon structure, and which may provide insight into the dynamics of confinement~\cite{Burkert:2018bqq}.

The mapping of the nucleon GPDs, and the detailed understanding of the spatial quark and gluon structure of the nucleon, have been widely recognized as key objectives of Nuclear Physics over the next decades. This requires a comprehensive program, combining results of measurements of a variety of processes in $eN$ scattering with structural information obtained from theoretical studies, as well as expected results from future lattice QCD calculations. In particular, GPDs can be accessed in the  lepto-production of real photons $lN \to lN\gamma$ through the Deeply Virtual Compton Scattering (DVCS) process corresponding to the scattering of a virtual photon into a real photon after interacting with a parton of the nucleon. At leading twist, DVCS accesses the 4 quark-helicity conserving GPDs $\{H_q,E_q,\widetilde{H}_q,\widetilde{E}_q\}$ defined for each quark-flavor $q\equiv\{u,d,s...\}$. They enter the cross section with combinations depending on the polarization states of the lepton beam and of the nucleon target, and are extracted from the modulation of experimental observables in terms of the out-of-plane angle $\phi$ between the leptonic and hadronic planes. The non-ambiguous extraction of GPDs from experimental data not only requires a large set of observables but also the separation of the different reaction amplitudes contributing to the $lN\gamma$ reaction. The combination of measurements with polarized lepton beams of opposite charges is an indisputable path towards such a separation~\cite{Diehl:2009:genova}.

This article investigates the opportunity and the benefits of the measurement of unpolarized and polarized Beam Charge Asymmetries (BCAs) for DVCS off the proton with the CLAS12 spectrometer~\cite{Burkert:2020akg} at the Thomas Jefferson National Accelerator Facility (JLab), using the existing highly polarized electron beam of the Continuous Electron Beam Accelerator Facility (CEBAF) and a future high-duty-cycle unpolarized and polarized positron beam~\cite{Grames:2019:loi,Accardi:2020swt}. The next section discusses the uniqueness of BCA observables for DVCS and their importance for the determination of GPDs. The experimental peculiarities of such measurements with CLAS12 are further addressed before presenting a detailed study of the impact of these potential data.
%
%----------------------------------------------------------------------------------------------------
%

\section{Beam charge asymmetries}
\label{sec:BCAs}

\subsection{Deeply Virtual Compton Scattering}
\label{sec:BCAs-1}

\begin{figure}[!ht]
\begin{center}
\includegraphics[width=0.99\linewidth]{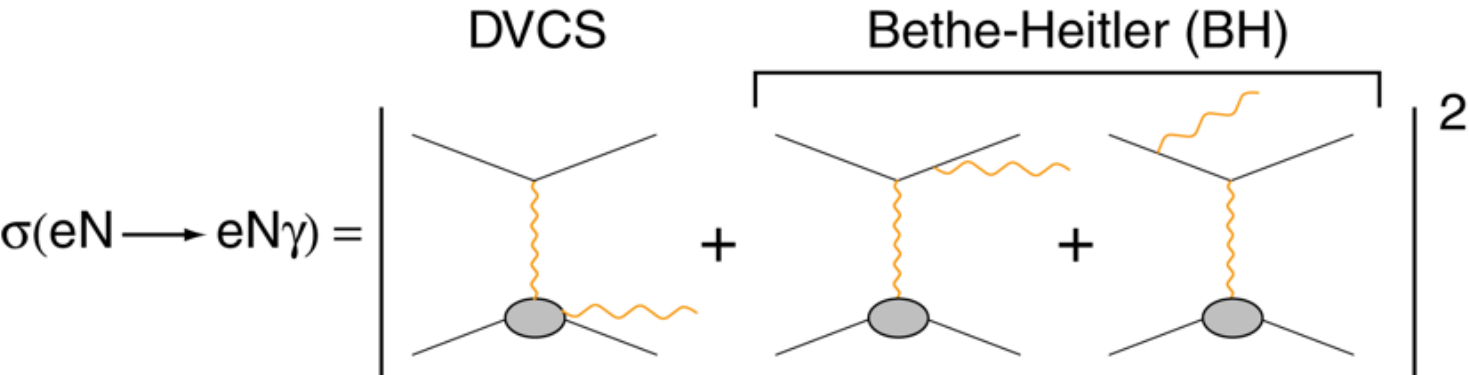}
\caption{Schematic of the lowest order QED amplitude of the electroproduction of real photons off nucleons.}
\label{EAgamma}
\end{center}
\end{figure}
Analogously to X-ray crystallography, the virtual \guillemotleft light \guillemotright produced by a lepton beam scatters on the partons to reveal the details of the internal structure of the proton (Fig.~\ref{EAgamma}). Because of this direct access to the parton structure, DVCS is the golden channel for the investigation of GPDs. This process competes with the known Bethe-Heitler (BH) reaction~\cite{Defurne:2017paw} where real photons are emitted from the initial or final leptons. The lepton beam charge ($e$) and polarization ($\lambda$) dependence of the $eN\gamma$ differential cross section off the proton is expressed~\cite{Diehl:2009:genova}
\begin{eqnarray}
d^5\sigma^{e}_{\lambda} & = & d^5\sigma_{BH} + d^5\sigma_{DVCS} + \lambda \, d^5\widetilde{\sigma}_{DVCS} \nonumber \\
 & + & e \, \left( d^5\sigma_{INT} + \lambda \, d^5\widetilde{\sigma}_{INT} \right) \label{eqXsec}
\end{eqnarray}
where
\begin{equation}
d^5\sigma^{e}_{\lambda} \equiv d^5\sigma^{e}_{\lambda} / dQ^2 dx_B dt d\phi_e d\phi
\end{equation}
with $Q^2$=-$q^2$ the four-momentum transfer of the virtual photon with energy $\omega$, $x_B$=$Q^2/2M\omega$ the Bjorken variable with the proton mass $M$, $t$=${(p-p')}^2$ the four-momentum transfer to the proton, and $\phi_e$ the azimuthal angle of the scattered electron. The index $INT$ denotes in Eq.~(\ref{eqXsec}) the $BH$-$DVCS$ quantum interference contribution to the cross section; $(d^5\sigma_{BH}$, $d^5\sigma_{DVCS}$, $d^5\sigma_{INT})$ represent the beam polarization independent contributions to the cross section, whereas ($d^5\widetilde{\sigma}_{DVCS}$, $d^5\widetilde{\sigma}_{INT}$) are the beam polarization dependent contributions. Polarized electron scattering only provides the experimental observables 
\begin{eqnarray}
\sigma^-_{0} & = & \frac{d^5\sigma^-_{+} + d^5\sigma^-_{-}}{2} \nonumber \\
 & = & d^5\sigma_{BH} + d^5\sigma_{DVCS} - d^5\sigma_{INT} \, , \label{eq:int00} \\
\Delta \sigma^-_{\lambda} & = & \frac{d^5\sigma^-_{+} - d^5\sigma^-_{-}}{2} = \lambda \, \left[ d^5\widetilde{\sigma}_{DVCS} -  d^5\widetilde{\sigma}_{INT} \right] 
\end{eqnarray}
which involve a combination of the unknwon $INT$ and $DVCS$ reaction amplitudes. The comparison between polarized electron and polarized positron reactions provides the additional observables
\begin{eqnarray}
\Delta \sigma_{0}^C & = & \frac{\sigma^+_{0} - \sigma^-_{0}}{2} = d^5\sigma_{INT} \\
\Delta \sigma_{\lambda}^C & = & \frac{\Delta\sigma^+_{\lambda} - {\Delta\sigma^-_{\lambda}} }{2} = \lambda \, d^5\widetilde{\sigma}_{INT} 
\end{eqnarray} 
which isolate the interference amplitude, and the observables
\begin{eqnarray}
\sigma_{0}^0 & = & \frac{\sigma^+_{0} + \sigma^-_{0}}{2} = d^5\sigma_{BH} + d^5\sigma_{DVCS} \\
\Delta \sigma_{\lambda}^0 & = & \frac{\Delta\sigma^+_{\lambda} + {\Delta\sigma^-_{\lambda}} }{2} = \lambda \, d^5\widetilde{\sigma}_{DVCS} 
\end{eqnarray} 
which select a $DVCS$ signal; the superscript $C$($0$) indicates the lepton beam charge (in)sensitive parts of the $eN\gamma$ cross section. Therefore, combining lepton beams of opposite polarities and different polarizations allows the separation of the 4 unknown $INT$ and $DVCS$ reaction amplitudes and permits an unambiguous access to combinations of GPDs. In the absence of such beams, the only possible approach to this separation is to take advantage of the different beam energy dependence of the $DVCS$ and $INT$ amplitudes. Recent results~\cite{Defurne:2017paw} have shown that this Rosen\-bluth-like separation cannot be performed without assumptions because of higher twists and higher $\alpha_s$-order contributions to the cross section. Positron beams in comparison to electron beams offer the most powerful experimental solution to this problem.

\subsection{Access to Generalized Parton Distributions}
\label{sec:BCAs-2}

\begin{figure}[!t]
\begin{center}
\includegraphics[width=0.330\textwidth]{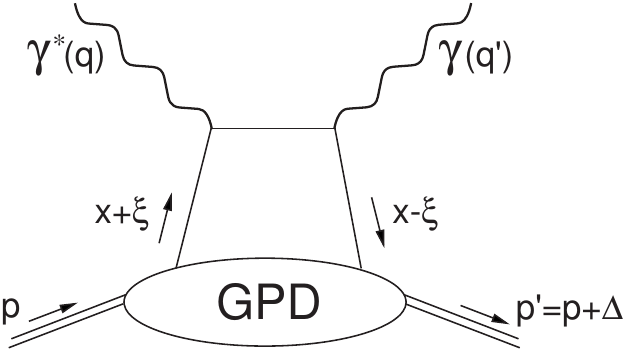}
\caption{Leading order and leading twist representation of the DVCS reaction amplitude ($+$ crossed term not shown) with the main kinematic parameters of the GPDs; $(q,q')$ are the quadri-momenta of the initial virtual and the final real photons; $(p,p')$ are the initial and final protons quadri-momenta.}
\label{GPD-diag}
\end{center}
\end{figure}
GPDs are universal non-perturbative objects entering the description of hard scattering processes. They are not a  positive-definite probability density but correspond to the amplitude for removing a parton carrying some longitudinal momentum fraction of the nucleon and restoring it with a different longitudinal momentum (Fig.~\ref{GPD-diag}). The skewness $\xi \simeq x_B / (2-x_B)$ measures the variation of the longitudinal momentum. In this process, the nucleon receives a four-momentum transfer $t$=${\mathbf{\Delta}}^2$ whose transverse component $\mathbf{{\Delta}_{\perp}}$ is the Fourier-conjugate of the transverse distance $\mathbf{r_{\perp}}$ between the active parton and the center-of-mass of spectator partons in the target~\cite{Burkardt:2007sc}. In the limit of zero-skewness ($\xi$=$0$), GPDs can be interpreted as the Fourier transform of the distribution in the transverse plane of partons with the longitudinal momentum fraction  $x$~\cite{Burkardt:2000uu,Ralston:2001xs,Diehl:2002he,Belitsky:2002ep}.

\begin{figure*}[t!]
\begin{center}
\includegraphics[width=0.995\textwidth]{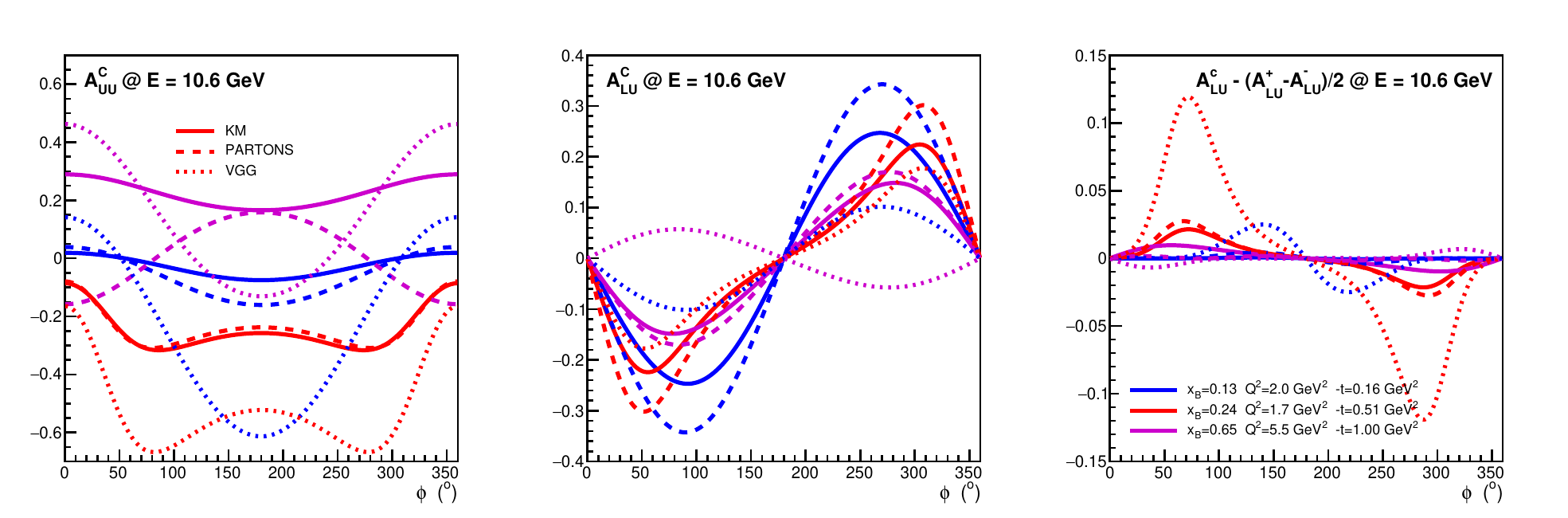}
\caption{Out-of-plane angular dependence of BCA observables for typical JLab kinematics from small to large $(x_B,Q^2,t)$. Calculations have been performed using three different GPD models: KM, PARTONS, and VGG as described in the text.}
\label{BCAssym}
\end{center}
\end{figure*}
GPDs enter the $eN\gamma$ cross section through Compton Form Factors (CFFs) ${\mathcal F}$ (with ${\mathcal F} \equiv \{ {\mathcal H}, {\mathcal E}, \widetilde{\mathcal H}, \widetilde{\mathcal E} \}$) defined as  
\begin{eqnarray}
{\mathcal F}(\xi,t) & = & {\mathcal P} \int_{0}^{1} dx \left[ \frac{1}{x-\xi} \pm \frac{1}{x+\xi} \right] F_+(x,\xi,t) \nonumber \\
& & \,\,\,\,\,\, - \, i \pi \, F_+(\xi,\xi,t) \label{eq:CFF}  
\end{eqnarray}
where ${\mathcal P}$ denotes the Cauchy's principal value integral, and  
\begin{equation}
F_+(x,\xi,t) = \sum_{q} \left(\frac{e_q}{e}\right)^2 {\left[ F^q(x,\xi,t) \mp F^q(-x,\xi,t) \right] }
\end{equation}
is the singlet GPD combination for the quark flavor $q$ where the upper sign holds for vector GPDs $(H^q,E^q)$ and the lower sign for axial vector GPDs $(\tilde{H}^q,\tilde{E}^q)$. Thus, the imaginary part of the CFF accesses GPDs along the diagonals $x$=$\pm\xi$ while the real part probes an integral over the initial longitudinal momentum of the partons of a convolution of GPDs and parton propagators. At leading twist and leading order, the CFF combinations entering the $DVCS$ and $INT$ contributions are
\begin{eqnarray}
{\mathcal F}_{DVCS} & = & 4(1-x_B) {\left( {\mathcal H} {\mathcal H}^{\star} + \widetilde{\mathcal H} \widetilde{\mathcal H}^{\star} \right)} - x_B^2 \frac{t}{4M^2} \widetilde{\mathcal E} \widetilde{\mathcal E}^{\star} \nonumber \\
& - & x_B^2 {\left( {\mathcal H} {\mathcal E}^{\star} + {\mathcal E} {\mathcal H}^{\star} + \widetilde{\mathcal H} \widetilde{\mathcal E}^{\star} + \widetilde{\mathcal E} \widetilde{\mathcal H}^{\star}\right)} \nonumber \\
& - & {\left( x_B^2 + (2-x_B)^2 \frac{t}{4M^2} \right)} {\mathcal E} \widetilde{\mathcal E}^{\star} \\
{\mathcal F}_{INT} & = & F_1 {\mathcal H} + \xi (F_1 + F_2) \widetilde{\mathcal H} - \frac{t}{4M^2} F_2 {\mathcal E} \, .
\end{eqnarray}
Separating the $INT$ contribution to the $eN\gamma$ cross section provides therefore a direct access to a linear combination of CFFs, as compared to the more involved bilinear combination of the $DVCS$ contributions. 

Analytical properties of the $DVCS$ amplitude at the Leading Order (LO) approximation lead to a dispersion relation between the real and imaginary parts of the CFFs~\cite{Anikin:2007yh,Diehl:2007jb,Polyakov:2007rv}
\begin{eqnarray}
\Re {\rm e} \left[ {\mathcal F}(\xi,t) \right] & \stackrel{\rm LO}{=} & D_{\mathcal F}(t) \\ 
& & \hspace*{-45pt} + \frac{1}{\pi}{\mathcal P}\int_{0}^{1} dx  \left
( \frac{1}{\xi-x}-\frac{1}{\xi+x}\right) \Im{\rm m} [{\mathcal F}(x,t)] \nonumber
\end{eqnarray}
where $D_{\mathcal F}(t)$ is the so-called $D$-term, a $t$-dependent subtraction constant. Originally introduced to restore the polynomiality property of vector GPDs, the $D$-term~\cite{Polyakov:1999gs} enters the parameterization of the non-forward matrix element of the Energy-Momentum Tensor which accesses the mechanical properties of the nucleon~\cite{Polyakov:2002yz,Burkert:2018bqq,Polyakov:2018zvc,Kumericki:2019ddg}. The independent experimental determination of the real and imaginary parts of the CFFs allows the extraction of the $D$-term, and is therefore a key feature for the understanding of nucleon dynamics.

\begin{figure*}[t!]
\begin{center}
\includegraphics[width=0.80\textwidth]{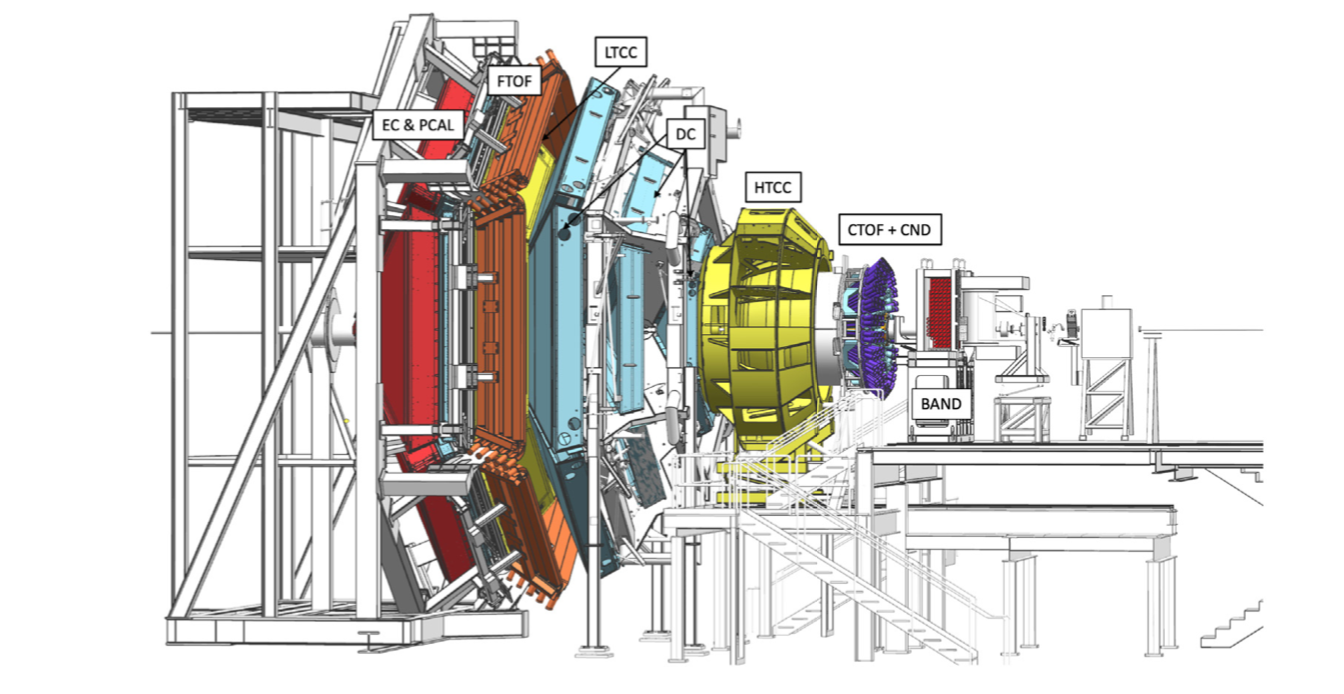}
\caption{The CLAS12 spectrometer in the Hall B at JLab~\cite{Burkert:2020akg}. Electron and positron beams come from the right and hit the target in the center of the solenoid magnet, which is at the core of the CD. It is largely hidden from view inside the HTCC \v{C}erenkov counter. Scattered particles are detected in the FD which comprises several different detectors immerged into the magnetic field of the torus magnet located after the CD.}
\label{clas12_fig}
\end{center} 
\end{figure*}

\subsection{Charge asymmetries}
\label{sec:BCAs-3}

Comparing polarized electron and positron beams, the unpolarized Beam Charge Asymmetry (BCA) $A^C_{UU}$ can be constructed following the expression 
\begin{eqnarray}
A^C_{UU} & = & \frac{(d^5\sigma^+_{+} + d^5\sigma^+_-) - ( d^5\sigma^-_+ + d^5\sigma^-_-)}{d^5\sigma^+_{+} + d^5\sigma^+_- + d^5\sigma^-_+ + d^5\sigma^-_-} \nonumber \\
& = & \frac{d^5 \sigma_{INT}}{d^5\sigma_{BH} + d^5 \sigma_{DVCS}}
\end{eqnarray}
where the superscript denotes the lepton beam charge and the subscript indicates its helicity. At leading twist, $A^C_{UU}$ is proportional to the CFF $\Re{\rm e} \left[{\mathcal F}_{INT}\right]$, and deviates from a linear dependence on the CFFs in the case of the non-dominance of the $BH$ amplitude with respect to the polarization-insensitive $DVCS$ amplitude. Similarly, the polarized BCA $A^C_{LU}$ can be constructed as
\begin{eqnarray}
A^C_{LU} & = & \frac{(d^5\sigma^+_{+} - d^5\sigma^+_-)/\lambda^+ - ( d^5\sigma^-_+ - d^5\sigma^-_-)/\lambda^-}{d^5\sigma^+_{+} + d^5\sigma^+_- + d^5\sigma^-_+ + d^5\sigma^-_-} \nonumber \\ 
& = & \frac{d^5 \widetilde{\sigma}_{INT}}{d^5 \sigma_{BH} + d^5 \sigma_{DVCS}} \label{eq:BCSA} 
\end{eqnarray}
which, at leading twist, is proportional to $\Im{\rm m} \left[{\mathcal F}_{INT}\right]$. As $A^C_{UU}$, $A^C_{LU}$ is a CFF signal affected by the same $BH$-non-dominance effects. At leading twist and in the $BH$-dominance hypothesis, $A^C_{LU}$ is opposite in sign to the Beam Spin Asymmetry (BSA) $A^-_{LU}$ measured with polarized electrons, and equal to the BSA $A^+_{LU}$ measured with polarized positrons. The relation 
\begin{equation}
A^C_{LU} = \frac{A^+_{LU} - A^-_{LU}}{2} \label{BHdom}
\end{equation}
can be viewed as a signature of the $BH$-dominance hypothesis and provides a handle on its validity. In the case of significant differences, the neutral BSA
\begin{eqnarray}
A^0_{LU} & = & \frac{(d^5\sigma^+_{+} - d^5\sigma^+_-)/\lambda^+ + ( d^5\sigma^-_+ - d^5\sigma^-_-)/\lambda^-}{d^5\sigma^+_{+} + d^5\sigma^+_- + d^5\sigma^-_+ + d^5\sigma^-_-} \nonumber \\ 
& = & \frac{d^5 \widetilde{\sigma}_{DVCS}}{d^5 \sigma_{BH} + d^5 \sigma_{DVCS}} \, ,
\end{eqnarray}
which is a twist-3 observable, allows us to distinguish the possible reasons for the breakdown of Eq.~(\ref{BHdom}). 

Pioneering DVCS measurements~\cite{Aaron:2009ac,Airapetian:2008aa,Airapetian:2009aa,Airapetian:2012mq} at HERA obtained  significant BCA signals in the gluon and sea-quark sectors. The COMPASS experiment operating high energy $\mu^{\pm}$ beams should release in the near future BCA data in the sea-quark region~\cite{Akhunzyanov:2018nut}, but there is no prospect in the valence-quark region. The advent of polarized positron beams at JLab~\cite{Grames:2019:loi} would allow measurements of BCA observables in this unexplored region typical of the JLab kinematic reach. Projections of BCA observables are shown in  Fig.~\ref{BCAssym} for selected kinematics at a 10.6~GeV beam energy. They are determined using the Belitsky-M\"uller (BM)  modeling of DVCS observables~\cite{Belitsky:2010jw} and either the Kumeri\v{c}ki-M{\"u}ller (KM)~\cite{Kumericki:2009uq}, the PARtonic Tomography Of Nucleon Software (PARTONS)~\cite{Berthou:2015oaw}, or a choice of Vanderhaeghen-Guichon-Guidal (VGG)~\cite{Vanderhaeghen:1999xj} CFFs. Asymmetry amplitudes are generally  very significant and sensitive to the CFF model, particularly the unpolarized BCA. The hypothesis of the BH-dominance appears as a kinematic- and model-dependent statement.  

%
%----------------------------------------------------------------------------------------------------
%

\section{Experimental configuration}
\label{sec:Expconfig}

The prospects for BCA measurements for the processes $\vv{e}^{\pm}p \to e^{\pm}p\gamma$ with the CLAS12 spectrometer (Fig.~\ref{clas12_fig}) \cite{Burkert:2020akg} is investigated further assuming alternating periods of electron and positron beams for a total duration of 80 days.

\subsection{Detector}
\label{sec:Expconfig-1}

The CLAS12 spectrometer combines a Central Detector (CD) surrounding the target and immersed into a high-intensity solenoidal feld, with a Forward Detector (FD) based on a six-coil torus magnet. This combination provides a large angular coverage and the ability to carry-out experiments at luminosities as high as 10$^{35}$~cm$^2 \cdot$s$^{-1}$. 

\begin{figure}[t!]
\begin{center}
\includegraphics[width=0.46\textwidth]{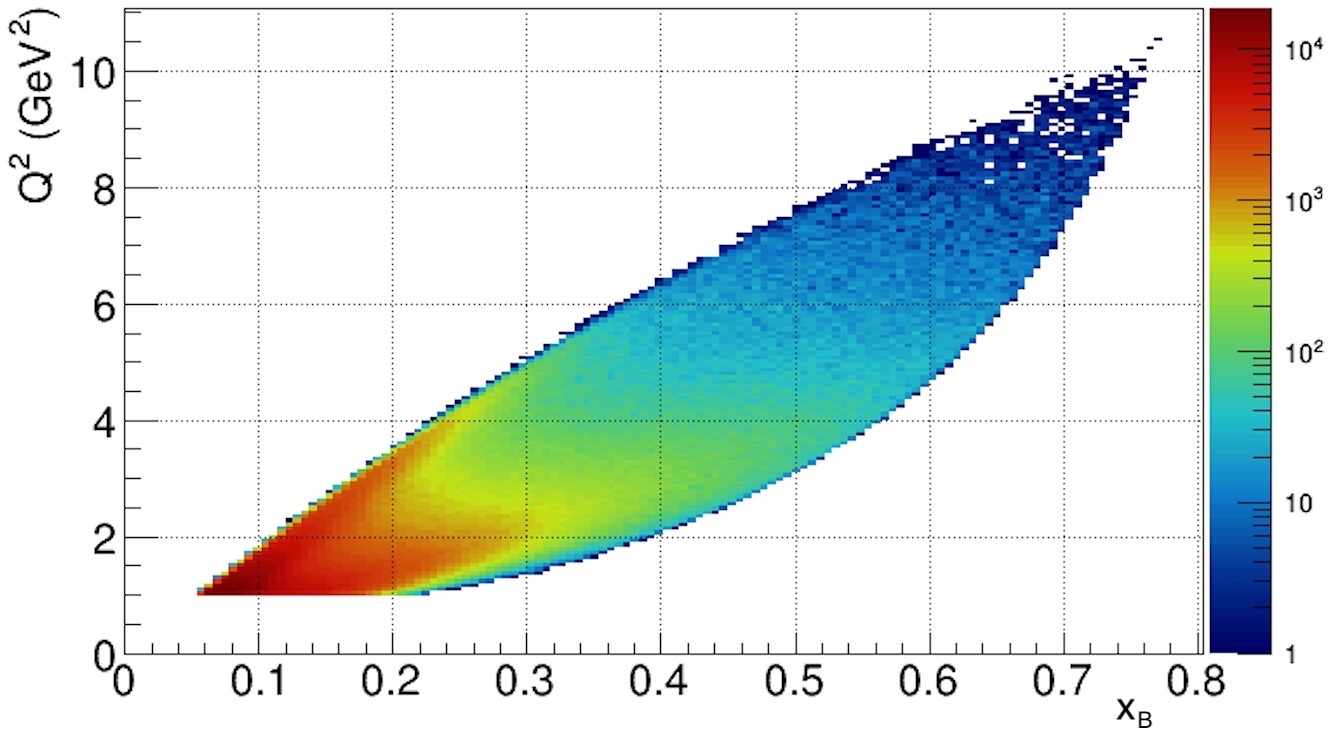}
\includegraphics[width=0.46\textwidth]{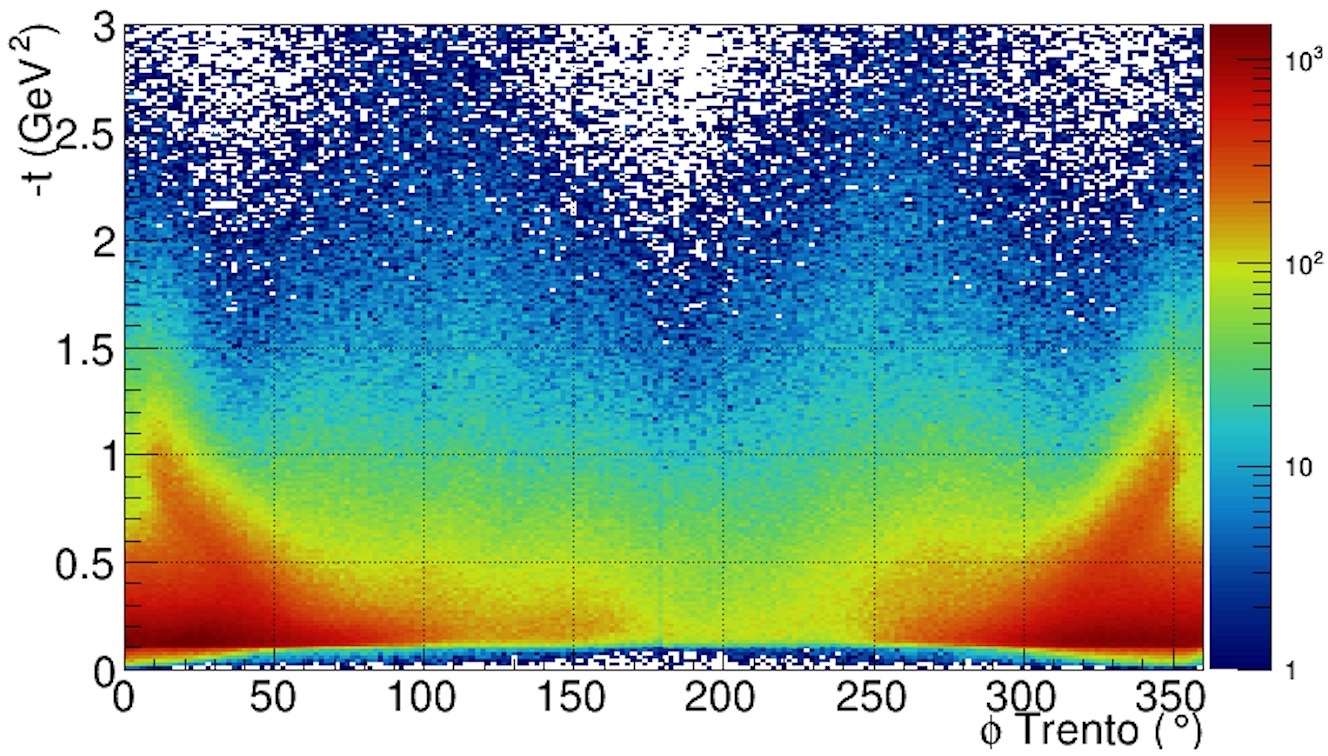}
\caption{Kinematic coverage of exclusive DVCS/BH events in $Q^2$ versus $x_B$ (top), and in -$t$ versus the azimuthal $\phi$-dependence (bottom), from a subset of electron beam data~\cite{Burkert:2020:pro}.}
\label{dvcs-kine}
\end{center}
\end{figure}

Scattered electrons or positrons are detected in the CLAS12 FDs including the high threshold \v{C}erenkov Counter (HTCC), the drift chamber trac\-king system (DC), the Forward Time-of-Flight system (FTOF) and the electromagnetic calorimeter (ECAL) which consists of the pre-shower calorimeter (PCAL) and the inner and outer parts of the electromagnetic calorimeter (EC). DVCS photons are measured in the CLAS12 ECAL that covers the polar angle range from about 5$^\circ$ to 35$^\circ$. Additionally, high energy photons are also detected in the Forward Tagger Calorimeter (FTCal), which spans the polar angle range of 2.5$^\circ$ to 4.5$^\circ$. Protons with momenta above 300~MeV/$c$ are detected mostly in the CLAS12 CD covering the 35$^{\circ}$-125$^{\circ}$ polar angle, but a significant fraction is also detected in the CLAS12 FD, especially those in the higher -$t$ range. The CD consists of several detector systems organized into successive layers around the reaction target: the Silicon Vertex Tracker (SVT), followed by the Barrel Micromesh Tracker (BMT), and completed with the Central Time-of-Flight (CTOF) and the Central Neutron Detector (CND) which read-out systems can be seen at the entrance of the solenoid magnet (Fig.~\ref{clas12_fig}). The Back Angle Neutron Detector (BAND) is further installed at the upstream end.  

The kinematic coverage of the DVCS process in the CLAS12 acceptance is shown in Fig.~\ref{dvcs-kine} for a subset of electron beam data~\cite{Burkert:2020:pro} at 10.6~GeV and a detector configuration similar to the positron configuration. When operating with positron beam, the experiment will use the standard Hall B beam line with the electrical diagnostics in reversed charge mode from operating the beam line and the experimental equipment with the electron beam. The experimental setup will be identical to the standard electron beam setup with both the Solenoid and the Torus magnets in reversed current mode from electron beam data taking. 

\subsection{Systematic uncertainties}
\label{sec:Expconfig-3}

The generation of polarized positron beams through the PEPPo method~\cite{Abbott:2016hyf} involves the interaction of the CEBAF primary electron beam with a thick high-$Z$ target to produce bremsstrahlung photons and subsequent $e^+e^-$-pairs for further acceleration into CEBAF. The properties of this secondary positron (as well as electron) beam are expected to differ from the primary electron beam, essentially by a 4-5 times larger emittance which may result in false asymmetries when comparing electron and positron beam data. Controlling these effects implies using an electron beam with properties similar to the secondary positron beam. Such a beam can be made of the secondary electrons simultaneously produced at the positron target and having the same properties as positrons in terms of $(x,y)$ profile and emittance. Comparing the detector response for primary and secondary electron beams, and/or acquiring physics data with both secondary electron and positron beams will resolve potential beam related false asymmetry issues.

\begin{figure}[ht]
\begin{center}
\includegraphics[width=1.0\columnwidth]{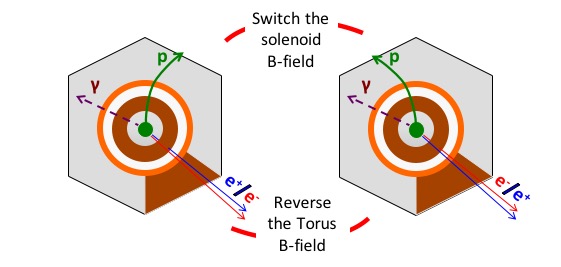}
\caption{The generic setup of the CLAS12 detector in Hall B viewed from upstream down the beam pipe. In this view the proton rotates in the opposite direction when operating beam of opposite polarities. When switching the solenoid field the electron and positron experience  different $\phi$-motions due to the opposite motion of electron and positrons (left). When the solenoid field is reversed the electrons and positrons get kicks in the opposite azimuthal directions as seen that positrons and electrons switch place in the forward detector (right).} \label{clas12-cartoon}
\end{center}
\end{figure} 
The general schematic of the CLAS12 spectrometer is shown in Fig.~\ref{clas12-cartoon} in a view  looking downstream along the beam line. For the scattered positrons and for the DVCS photons the detector looks identical to the situation when electrons are scattered off protons and the magnetic fields in both magnets are reversed. This is not the case for the recoil protons, which are bent in the solenoid field in the opposite direction compared to the electron scattering case. This could result in systematic effects due to different track  reconstruction efficiencies and effective solid angles leading to corrections of the raw yield asymmetries. Considering the unpolarized BCA, the effect of these corrections can be expressed as   
\begin{equation}
A^C_{UU} = \frac{\left( 1 + \eta_c \right) \, \mathcal{Y}^C_{UU} - \eta_c}{1 + \eta_c - \eta_c \, \mathcal{Y}^C_{UU}} \label{ACuu:YCUU}
\end{equation}
where the raw unpolarized Yield Charge Asymmetry (YCA) is defined as
\begin{equation}
\mathcal{Y}^C_{UU} = \frac{(Y^+_+ + Y^+_-) - (Y^-_+ + Y^-_-)}{Y^+_+ + Y^+_- + Y^-_+ + Y^-_-}
\end{equation}
from the beam charge- and spin-dependent normalized yield 
\begin{equation}
Y^e_{\lambda} = \frac{N^e_{\lambda}}{Q^e_{\lambda}} \, \frac{1}{\epsilon^e} \, .
\end{equation}
In the latter expression, $N^e_{\lambda}$ is the number of events in the solid angle $d^5 \Omega^e$, $Q^e_{\lambda}$ is the beam spin-dependent accumulated charge, and $\epsilon^e$ is the beam spin-independent detector efficiency. The correction factor $\eta_c$ is defined as  
\begin{equation}
\eta_c = C_{\Omega} - C_{\epsilon} - 2 \, C_{\epsilon} C_{\Omega}
\end{equation}
where
\begin{eqnarray}
C_{\epsilon} & = & \frac{1}{2} \, \frac{\epsilon^+ - \epsilon^-}{\epsilon^+} \\
C_{\Omega}   & = & \frac{1}{2} \, \frac{\Delta \Omega^+ - \Delta \Omega^-}{\Delta \Omega^-}
\end{eqnarray}
with
\begin{equation}
\Delta \Omega^{\pm} = \int_{Bin} dQ^2 \, dx_B \, dt \, d\phi_e \, d\phi \, ,
\end{equation}
which quantifies the detector response differences between electron and positron data taking. Similarly, the polarized BCA 
is obtained from the raw polarized YCA
\begin{equation}
\mathcal{Y}^C_{LU} = \frac{(Y^+_+ - Y^+_-)/\lambda^+ - (Y^-_+ - Y^-_-)/\lambda^-}{Y^+_+ + Y^+_- + Y^-_+ + Y^-_-}
\end{equation}
according to the expression
\begin{figure}[!t]
\begin{center}
\includegraphics[width=1.0\columnwidth]{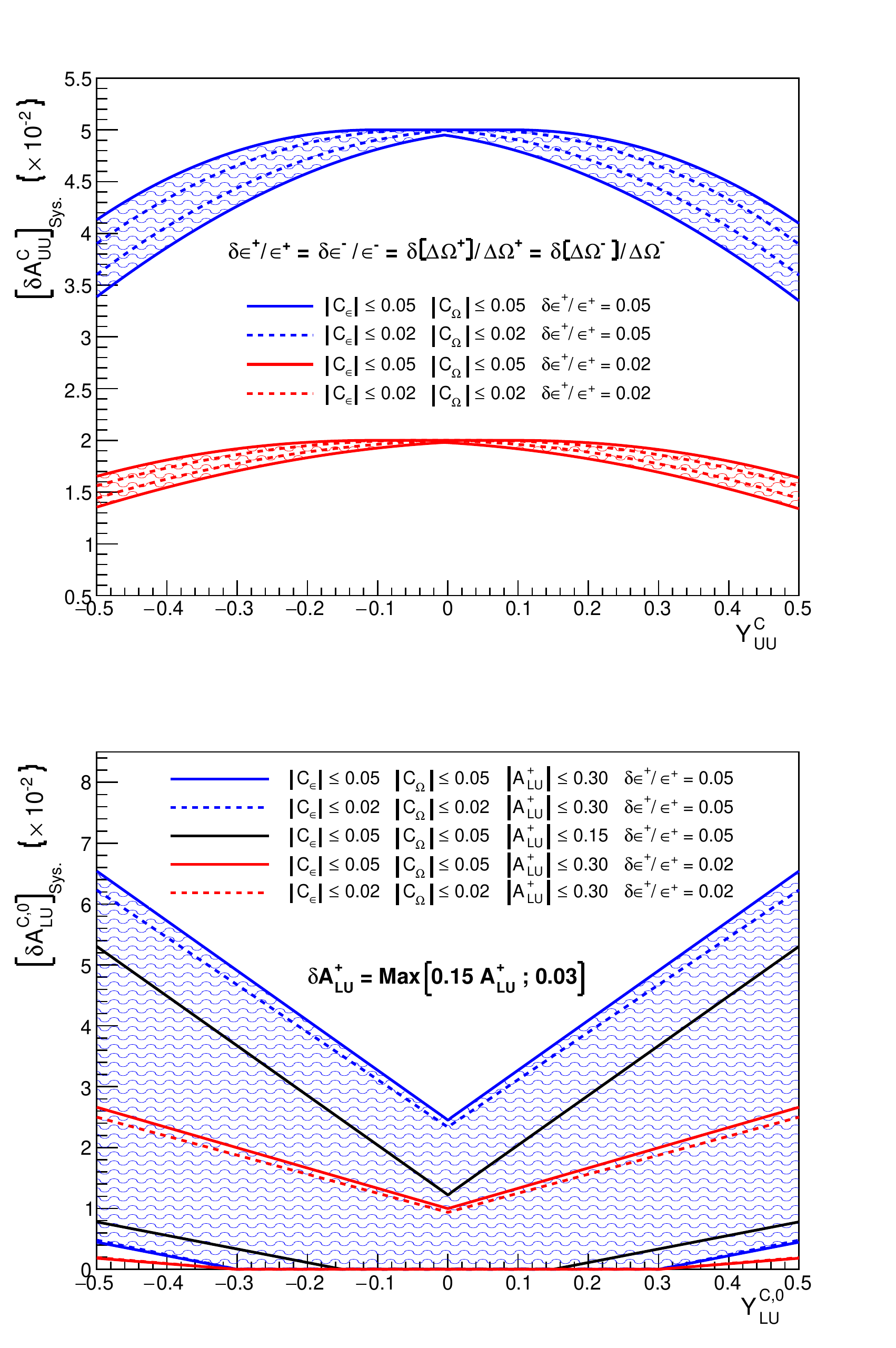}
\caption{Magnitude of BCA systematics induced by the detector response differences between positron and electron beams operation. The shaded areas indicate the possible values of BCA systematics assuming that unpolarized and polarized YCAs are comprised within $\pm$50\% and scanning any possible values of the correction factors $(C_{\epsilon},C_{\Omega})$ within $\pm$0.05 (full lines) or $\pm$0.02 (dashed lines). The different colors distinguish the values of the uncertainties on the detector efficiency and solid angle which are assumed to be identical for the sake of simplicity. The effect of the magnitude of the positron BSA is further shown on the bottom plot by comparing the areas delimited by the blue and black full lines. Only the boundaries of the shaded area for the $\delta \epsilon^+ / \epsilon^+$=0.02 case are shown for the  polarized BCAs.} 
\label{SysBCA}
\end{center}
\end{figure}
\begin{equation}
A^C_{LU} = \mathcal{Y}^C_{LU} + \eta_c \left( 1+ A^C_{UU} \right) \left[ \mathcal{Y}^C_{LU} - A^+_{LU} \right] \label{ACLu:YCLU} \, ,
\end{equation}
and the neutral BSA follows from the raw neutral Yield Spin Asymmetry (YSA) 
\begin{equation}
\mathcal{Y}^0_{LU} = \frac{(Y^+_+ - Y^+_-)/\lambda^+ + (Y^-_+ - Y^-_-)/\lambda^-}{Y^+_+ + Y^+_- + Y^-_+ + Y^-_-}
\end{equation}
according to
\begin{equation}
A^0_{LU} = \mathcal{Y}^0_{LU} + \eta_c \left( 1+ A^C_{UU} \right) \left[ \mathcal{Y}^0_{LU} - A^+_{LU} \right] \label{A0Lu:Y0LU} \, .
\end{equation}
The corrections to the raw YCAs and YSA can be small to sizeable but are exactly calculable once efficiency and solid angle effects are known. However, the precision on these corrections enters directly the systematic error on the experimental observables, the effect of which can be evaluated according to the expressions
\begin{eqnarray}
{\left( \delta \left[ A^C_{UU} \right]_{Sys.} \right)}^2 & = & {\left[ \frac{dA^C_{UU}}{dC_{\epsilon}} \, \delta C_{\epsilon} \right]}^2 + {\left[ \frac{dA^C_{UU}}{dC_{\Omega}} \, \delta C_{\Omega} \right]}^2 \\
{\left( \delta \left[ A^{C,0}_{LU} \right]_{Sys.} \right)}^2 & = & {\left[ \frac{dA^{C,0}_{LU}}{dC_{\epsilon}} \, \delta C_{\epsilon} \right]}^2 + {\left[ \frac{dA^{C,0}_{LU}}{dC_{\Omega}} \, \delta C_{\Omega} \right]}^2 \nonumber \\
 & + & {\left[ \eta_c \left( 1+ A^C_{UU} \right) \, \delta A^+_{LU} \right]}^2 
\end{eqnarray}
where we note the additional contribution from the positron BSA in the polarized case. Such an evaluation is performed in Fig.~\ref{SysBCA} which shows the envelope of systematic uncertainties depending on the raw YCAs and YSA and on the $(\delta C_{\epsilon},\delta C_{\Omega})$ corrections accuracy determined assuming fixed values - 0.05 (full lines) and 0.02 (dashed lines) - of the efficiency and solid angle precision. While the magnitude of systematics obviously depends on the magnitude of $(\delta C_{\epsilon},\delta C_{\Omega})$, the envelope is only weakly sensitive to the exact value $(C_{\epsilon},C_{\Omega})$ of the corrections. The sensitivity to the raw YCAs and YSA suggests stronger effects on the $\phi$-dependence of polarized observables than on unpolarized ones. For instance, considering the neutral BSA which is predicted to be zero in the twist-2 approximation, systematic effects may account for up to 0.03 depending essentially on the magnitude of the positron BSA.

From an experimental point of view, the measurement of a known process simultaneously to DVCS data taking will procure the calibration and the monitoring of these effects within a self-consistent calibrated simulation scheme. The $e^{\pm}p \to e^{\pm}p$ elastic scattering at small $Q^2$, {\it i.e.} in a region where 2-photon effects are very small~\cite{Afanasev:2017gsk}, is an approriate candidate for this purpose. 

\subsection{Projected data}
\label{sec:ProjDat}

\begin{figure}[!t]
\begin{center}
\includegraphics[width=0.342\textwidth]{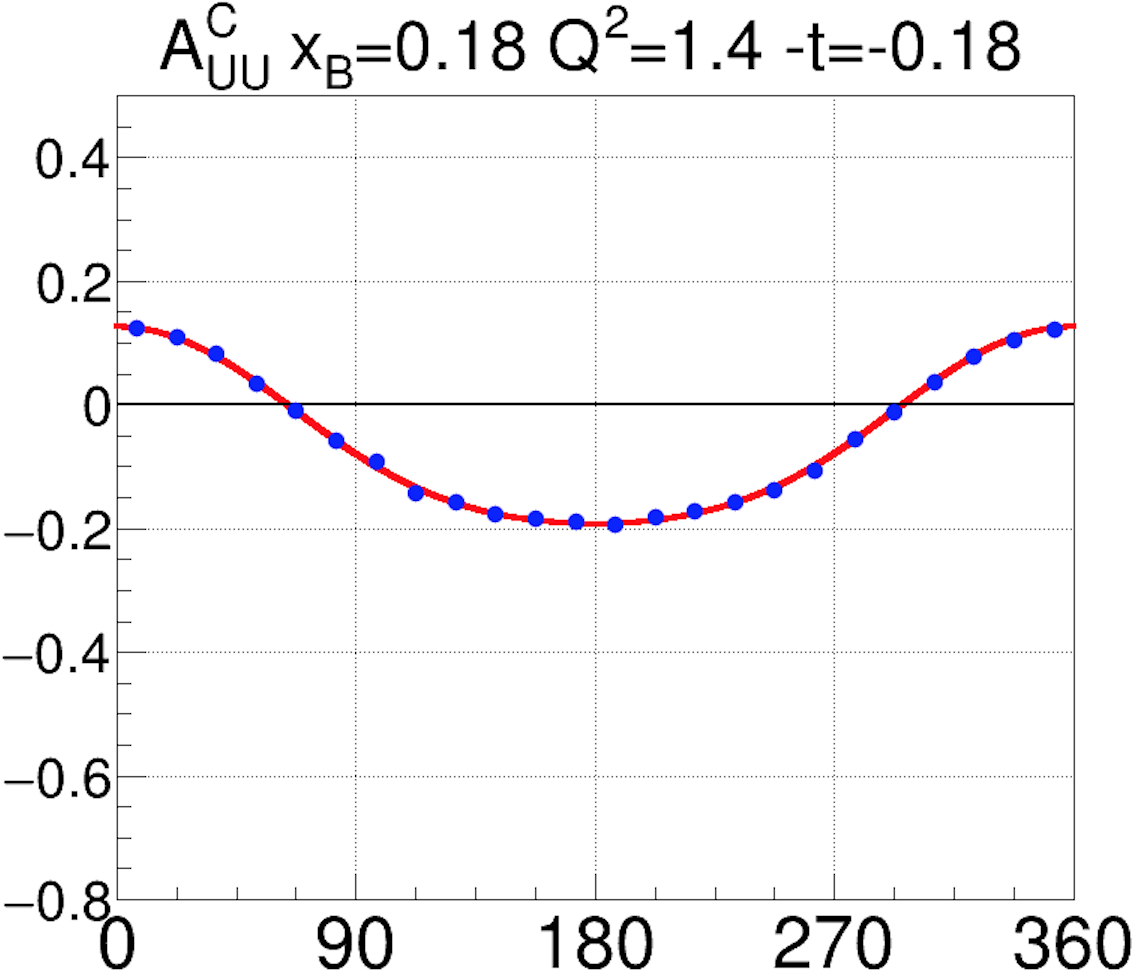}
\includegraphics[width=0.342\textwidth]{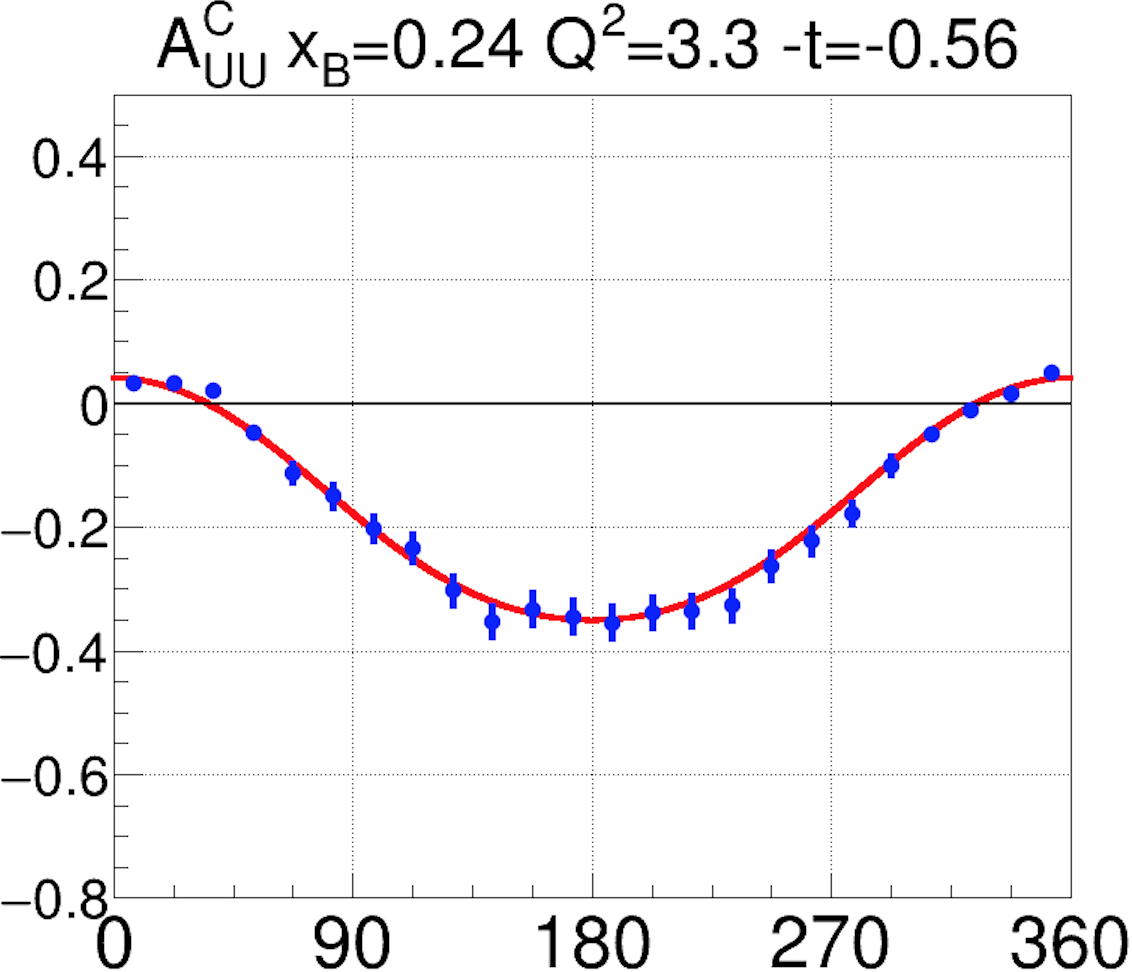}
\includegraphics[width=0.342\textwidth]{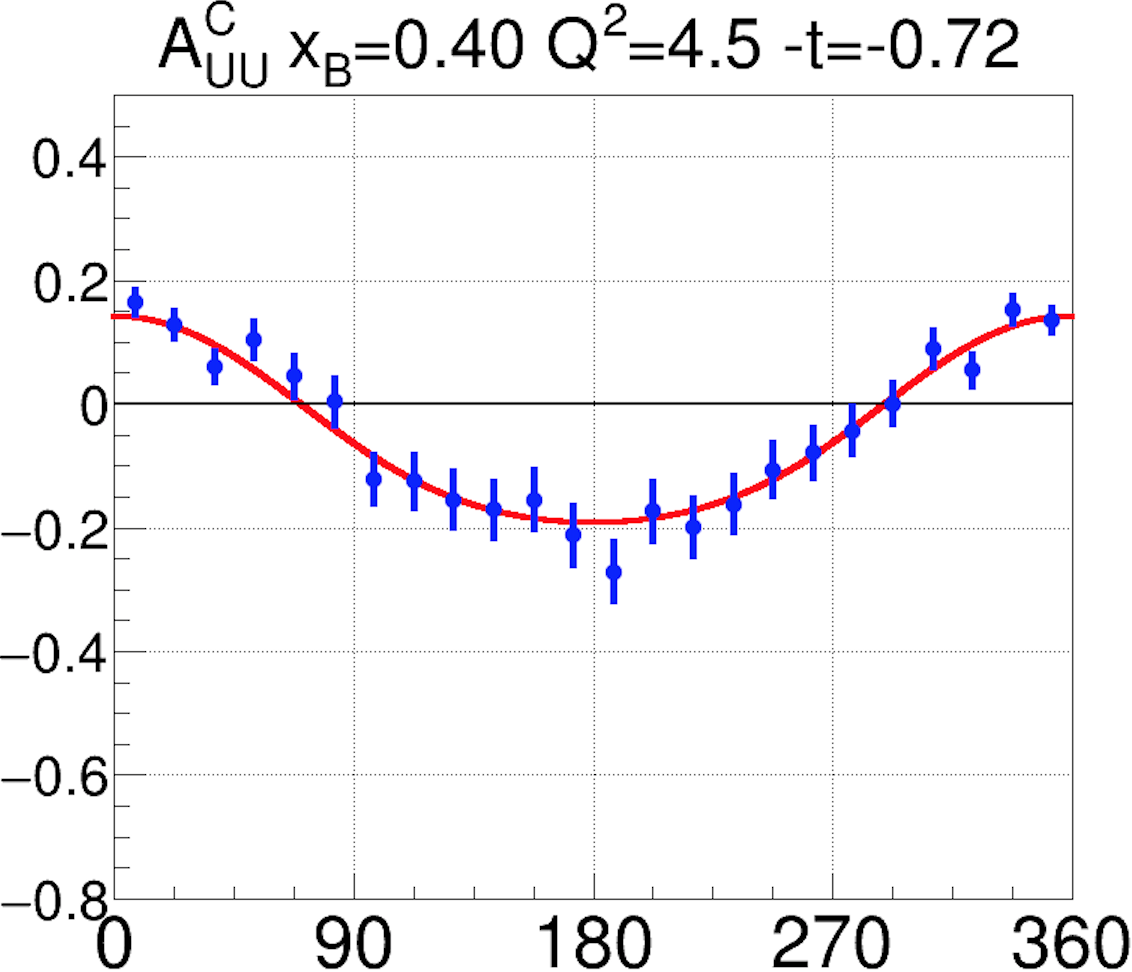}
\includegraphics[width=0.342\textwidth]{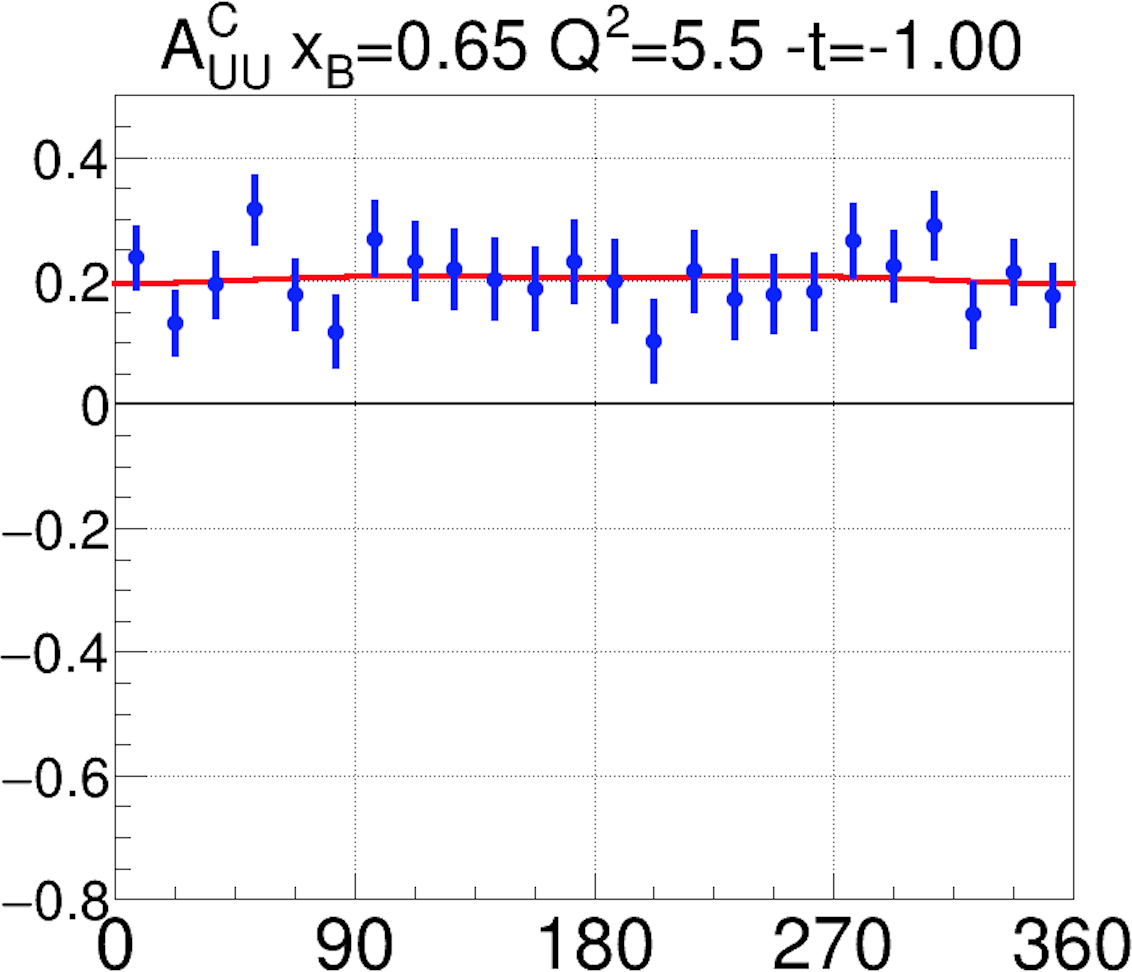}
\caption{Out-of-plane $\phi$-distribution of a subset of projected $A^C_{UU}$ data for selected bins~\cite{Burkert:2020:pro}; blue points correspond to projected data smeared according to their statistical error; red lines are the model prediction used to generate experimental observables.}
\label{projdat-UU}
\end{center}
\end{figure}

\begin{figure}[!h]
\begin{center}
\includegraphics[width=0.342\textwidth]{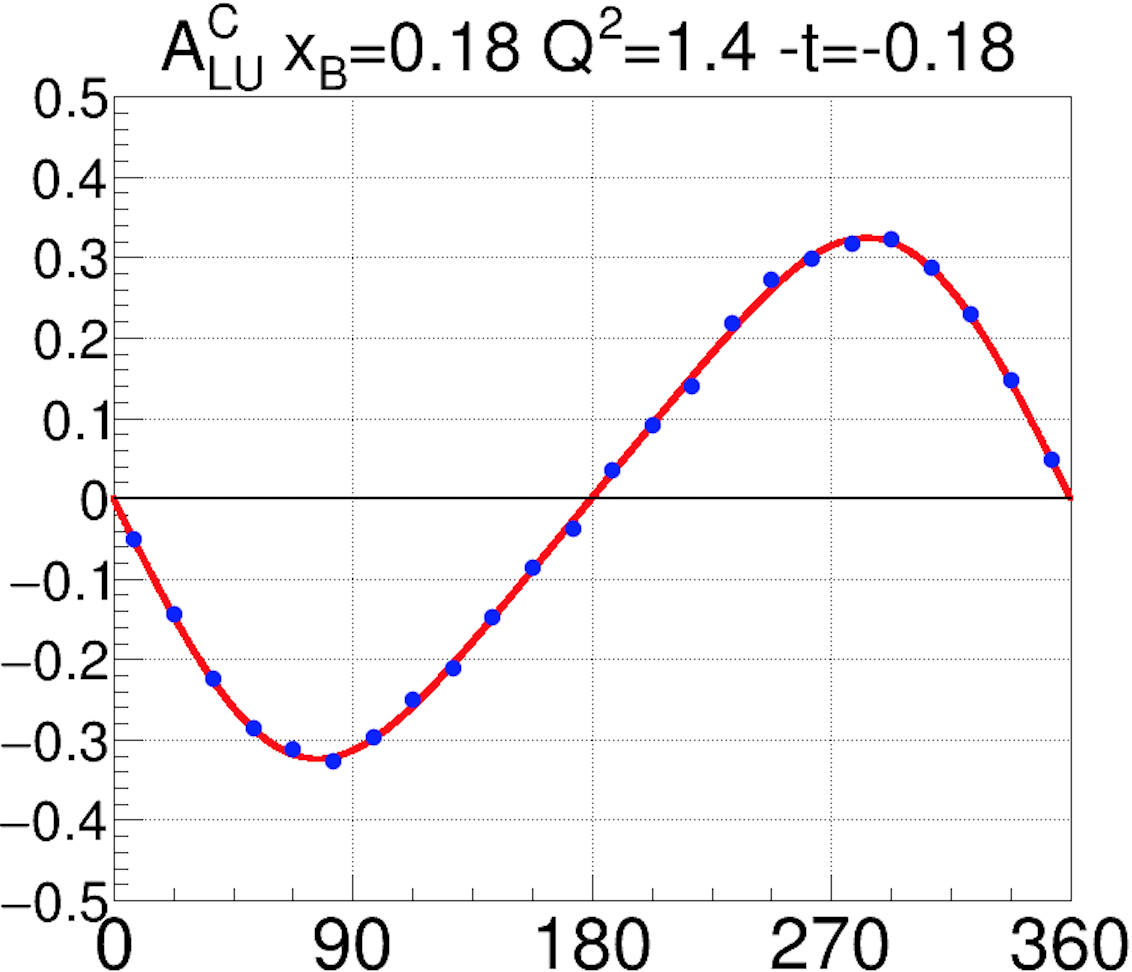}
\includegraphics[width=0.342\textwidth]{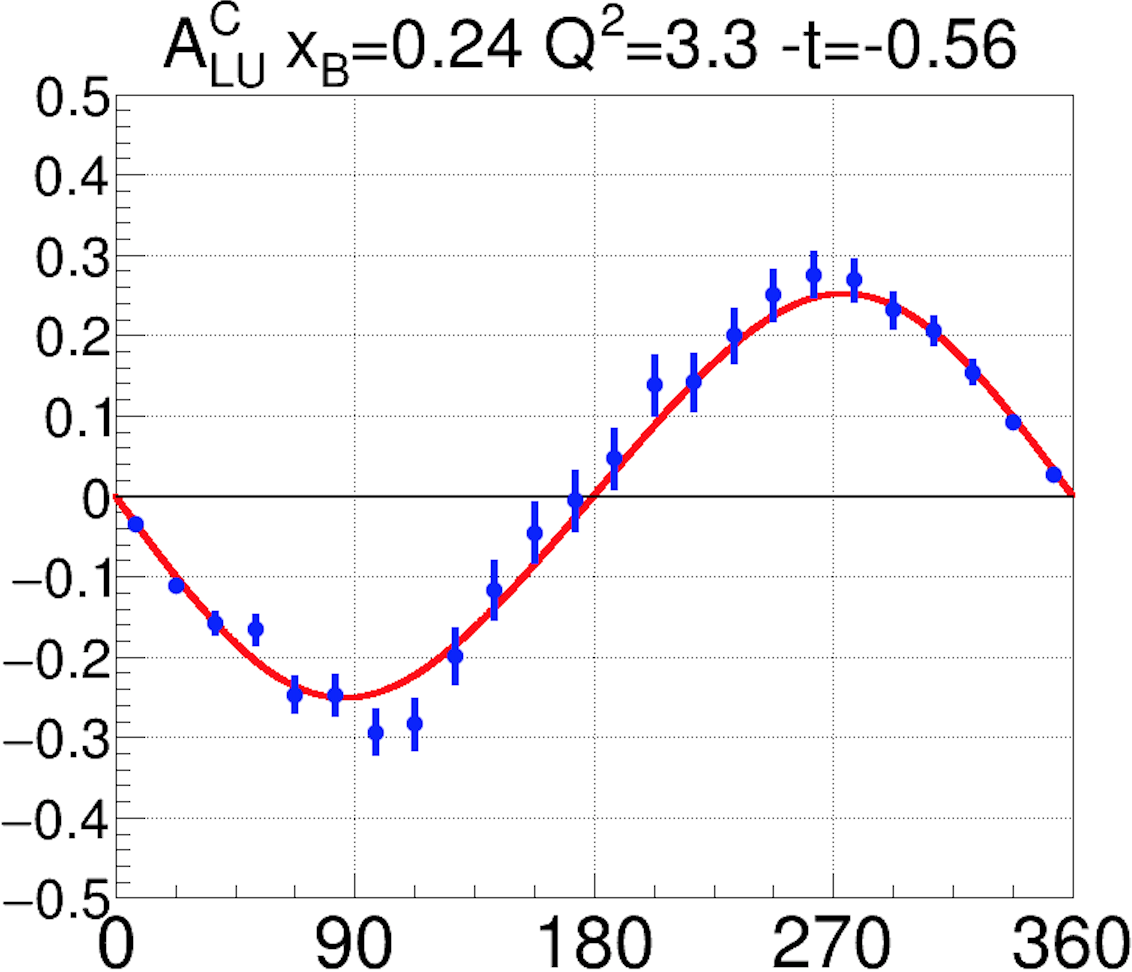}
\includegraphics[width=0.342\textwidth]{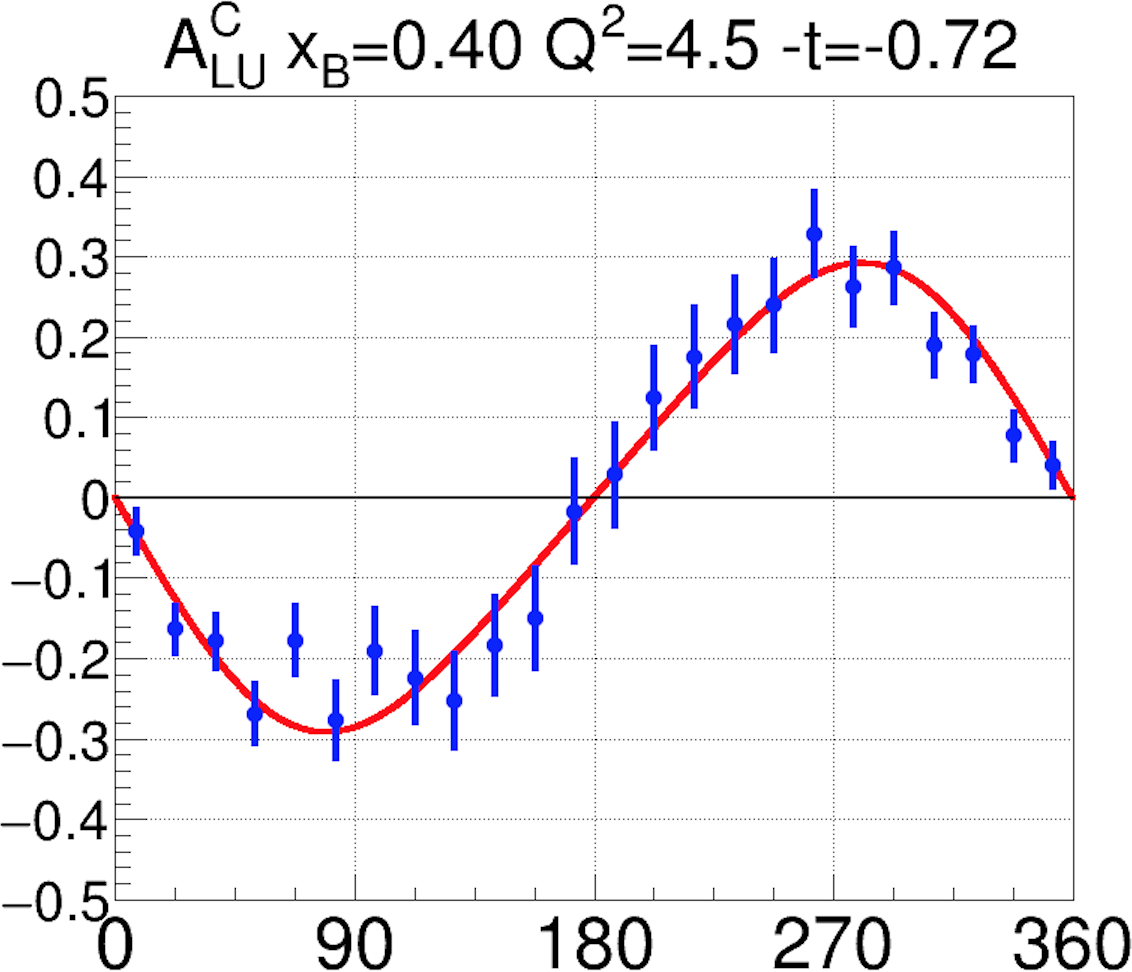}
\includegraphics[width=0.342\textwidth]{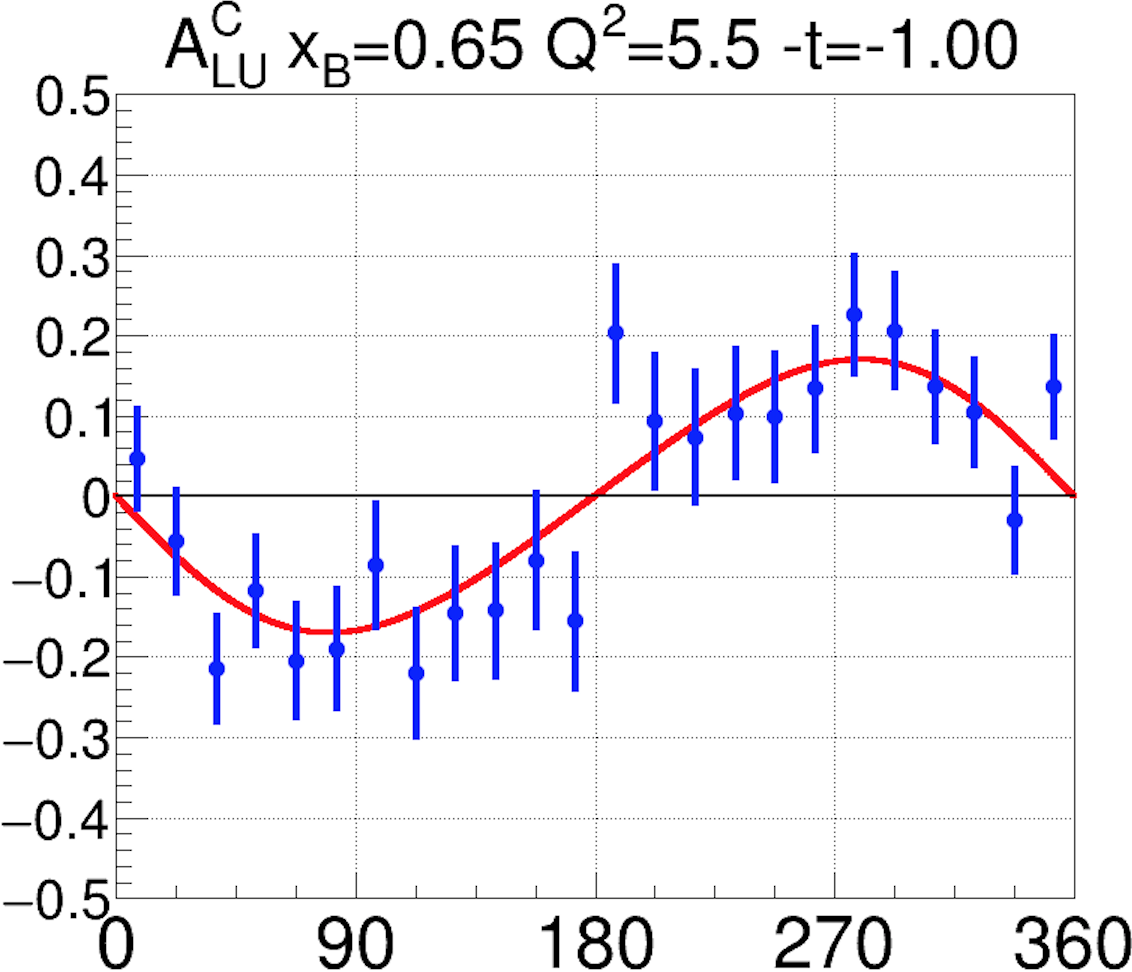}
\caption{Out-of-plane $\phi$-distribution of a subset of projected $A^C_{LU}$ data for selected bins~\cite{Burkert:2020:pro}; blue points correspond to projected data smeared according to their statistical error; red lines are the model prediction used to generate experimental observables.}
\label{projdat-LU}
\end{center}
\end{figure}

Thanks to the large kinematic coverage and the luminosity capabilities of the CLAS12 spectrometer, BCA observables could be efficiently investigated in the valence quark region at JLab. The performance of a potential measurement is hereafter evaluated for a beam energy of 10.6~GeV and the BM modeling of the $ep\gamma$ cross section. Statistical errors assume 80 days of running shared between electron and positron beams, and a luminosity of 0.6$\times 10^{35}$~cm$^{-2}\cdot$s$^{-1}$. A subset of projected data is shown on  Fig.~\ref{projdat-UU}-\ref{projdat-LU} for typical kinematics, exhibiting signals of different magnitude and shape. BCA observables are particularly well-defined at small $(x_B,Q^2,\vert t \vert)$. Following the kinematic dependence of the cross section, the statistical precision of observables degrades as any of the kinematic parameters increases. Nevertheless, BCA observables retain a significant selective power until the upper end of the physics domain accessible at CLAS12. 

%
%----------------------------------------------------------------------------------------------------
%

\section{Impact of positron measurements}
\label{sec:Impact}

The importance of BCA observables for the extraction of CFFs has been stressed numerous times in the literature (see among others~\cite{Diehl:2003ny} or~\cite{BELITSKY20051}). At twist-2 this problem can be seen as the determination of 8 unknown quantities (4 $\Re{\rm e} \left[{\mathcal F}\right]$ and 4 $\Im{\rm m} \left[{\mathcal F}\right]$) from a non-linear system of coupled equations~\cite{Guidal:2008ie}, which requires a minimum of 8 independent experimental observables with different sensitivities to the unknown quantities. Dispersion relations, sum rules, and QCD evolution bring correlations between CFF and links with elastic and deep inelastic experimental data, but the problem is generally complex and requires a large set of experimental observables. 

The methods for the extraction of CFFs from observables can be classified in two generic groups: GPD-model independent (local fit)~\cite{Guidal:2008ie,Kumericki:2011rz} and dependent (global fit)~\cite{Kumericki:2009uq,Moutarde:2019tqa} methods. Both methods are still depending on the cross section model (leading twist, target mass corrections, higher twist, NLO corrections...) and on further fitting hypotheses like the number of CFFs to be extracted from data. In an attempt at a quantitative evaluation of the impact of positron measurements, a local fit approach has been developed to extract the ${\mathcal H}$ and $\widetilde{\mathcal H}$ CFFs. Because of the different hypotheses quoted previously, this evaluation is inevitably model-dependent.


\begin{figure*}[!t]
\begin{center}
\includegraphics[width=0.85\textwidth]{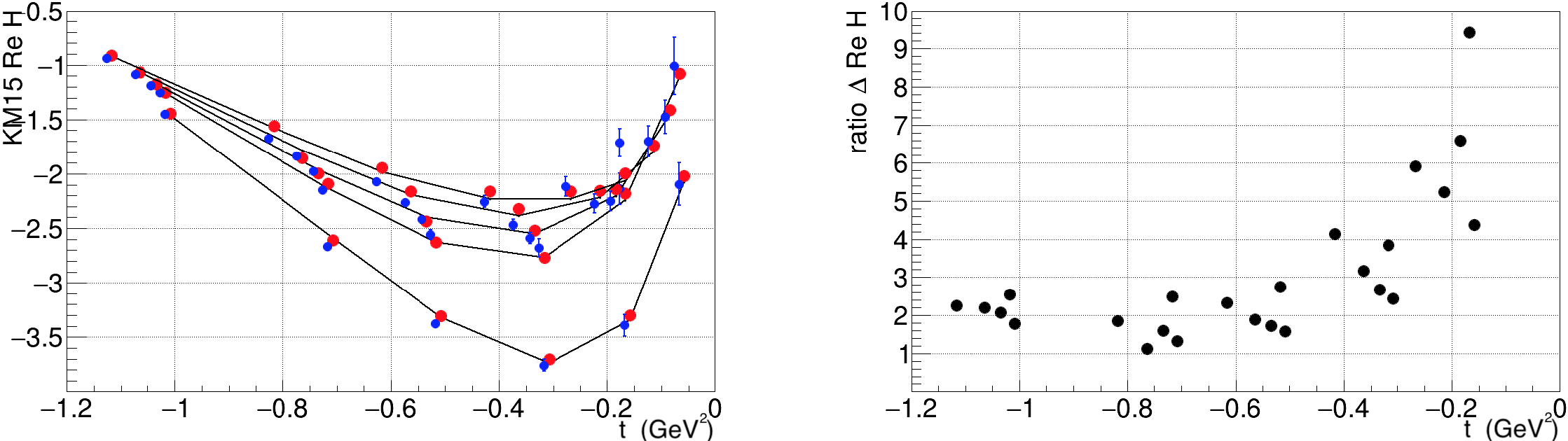}
\includegraphics[width=0.85\textwidth]{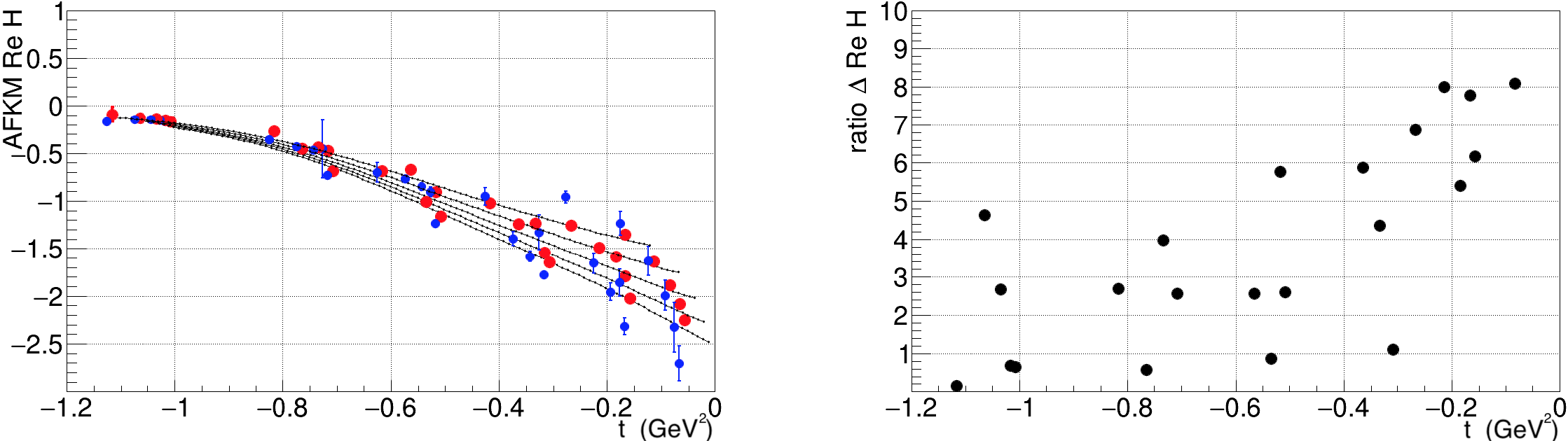}
\caption{Impact of the positron data on the extraction of $\Re {\rm e} [\mathcal{H}]$ for the KM15 (top)~\cite{Kumericki:2009uq} and AFKM (bottom)~\cite{Aschenauer:2013hhw} CFF parameterizations. On the left panels are the projections of extracted $\Re {\rm e} [\mathcal{H}]$ without (blue points) and with (red points) positron data in comparison to the model value (line); on the right panel are the ratios of the error on the extracted $\Re {\rm e} [\mathcal{H}]$ with positron data with the error for electron data only. The blue points are slightly shifted horizontally for visual clarity.}
\label{impact}
\end{center}
\end{figure*}
Within the present study, the benefit of positron measurements is quantified with respect to the CFF extraction from a fit of already approved CLAS12 DVCS measurements on the proton with an electron beam ($d^5\sigma^-_{UU}$, $A^-_{LU}$, $A^-_{UL}$, $A^-_{LL}$) with or without BCA data. In absence of completed analysis or existence of experimental data, we consider the data taking parameters of these approved experiments to determine statistical errors. Without impact on the extraction of $\mathcal{H}$, the transversely polarized target is not considered in this study. Experimental observables are determined over the full kinematic range of CLAS12 for a 10.6~GeV beam energy using the BM modeling of the cross section combined with different CFF parameterizations~\cite{Kumericki:2009uq,Aschenauer:2013hhw}. Individual observables are randomly smeared with the projected statistical uncertainties, and systematically shifted with the projected systematic uncertainties before CFF fitting. The CFF $\mathcal{H}$ and $\widetilde{\mathcal{H}}$ are then simultaneously extracted from projected data using a fitting procedure which assumes the model values for the non-fitted CFFs. The results are reported on Fig.~\ref{impact} for the full set of kinematics accessible with CLAS12, using the 2 CFF parameterizations. The left panel shows the model $\Re {\rm e} [\mathcal{H}]$ as a function of $-t$ for different $(x_B,Q^2)$ bins (line), together with the extracted values without (blue points) and with (red points) ($A^C_{UU},A^C_{LU}$) positron data. The right panel shows the ratios of CFF uncertainties. The impact of positron data is found to be particularly strong at small -$t$ where they can decrease uncertainties on $\Re {\rm e} [\mathcal{H}]$ by a large fraction, the magnitude of which depends on the CFFs parameterization. The electron-data-only scenario tends to provide different values from the model ones. By procu\-ring a pure interference signal, positron data reduce the correlations between CFFs and allow the fitting procedure to  recover the input model value. 

%
%----------------------------------------------------------------------------------------------------
%

\section{Conclusion}
\label{sec:conc}

The existence of a high-duty-cycle polarized positron beam at JLab would permit the investigation of the proton structure with new physics observables, complementary to those measured with a polarized electron beam. In particular, the comparison of electron and positron response of the proton through BCAs in the DVCS channel provides a pure $BH$-$DVCS$ interference signal. This ensures an accurate extraction of the $\Re {\rm e} [\mathcal{H}]$ CFF, of importance for the understanding of the dynamics underlying nucleon structure. Within a reasonable amount of beam time and using the CLAS12 spectrometer, a large $(x_B,Q^2,\vert t \vert)$ phase space in the valence quark region would be efficiently measured.

The present DVCS project is part of a wide community effort for the development of polarized positron beams at JLab~\cite{Accardi:2020swt}. It was approved by the 48$^{\mathrm{th}}$ JLab Program Advisory Committee assuming that the final positron source would reach the expected performances. The Conceptual Design Report about the implementation of polarized and unpolarized positron beams at CEBAF should be delivered by the end of 2022, opening the perspective of performing positron experiments at JLab in the second half of the 2020's.

%
%----------------------------------------------------------------------------------------------------
%

\begin{acknowledgements}

This article is part of a project that has received funding from the European Union's Horizon 2020 research and innovation program under agreement STRONG - 2020 - No~824093. It is based upon work supported by the U.S. Department of Energy, Office of Science, Office of Nuclear Physics under contract DE-AC05-06OR23177.

\end{acknowledgements}

%
%----------------------------------------------------------------------------------------------------
%

\bibliographystyle{spphys}       % APS-like style for physics

\bibliography{pDVCS-BCA}

%
%----------------------------------------------------------------------------------------------------
%

\end{document}